\documentclass[11pt,a4paper, onecolumn]{article}

\usepackage{url,colortbl}
\usepackage{xcolor}
\usepackage{bm,amsmath,amssymb}
\usepackage[]{graphicx}
\usepackage[margin=2cm]{geometry}
\usepackage[ruled,linesnumbered]{algorithm2e}
\usepackage{todonotes}
\graphicspath{{./Images/}}
\usepackage{authblk}
\usepackage{xr}
\usepackage{listings} 
\usepackage{dsfont}
\usepackage{appendix}
\usepackage[font=small]{caption}
\usepackage{setspace}
\usepackage{cite}
\usepackage{mathtools}
\usepackage{bigints}
\usepackage{mhchem}

\begin{document}

\setlength{\parindent}{0pt}
\setlength{\parskip}{1em}

\doublespacing

\onecolumn

\renewcommand{\abstractname}{}
\title{An energy-based mathematical model of actin-driven \\ protrusions in eukaryotic chemotaxis}
\author[1]{Samuel W.S. Johnson\thanks{Corresponding author: samuel.johnson@chch.ox.ac.uk}}
\author[2]{Maddy Parsons}
\newcommand\CoAuthorMark{\footnotemark[\arabic{footnote}]}
\author[1]{Ruth E. Baker\thanks{These authors contributed equally.}}
\author[1]{Philip K. Maini\protect\CoAuthorMark}
\affil[1]{Wolfson Centre for Mathematical Biology, Mathematical Institute, University of Oxford, Oxford, United Kingdom}
\affil[2]{Randall Centre for Cell and Molecular Biophysics, King’s College London, London, United Kingdom}
\date{}
\maketitle

\linespread{2}\selectfont


\vspace{-36pt}

\begin{abstract}
    \noindent In eukaryotic cell chemotaxis, cells extend and retract transient actin-driven protrusions at their membrane that facilitate both the detection of external chemical gradients and directional movement via the formation of focal adhesions with the extracellular matrix. Although extensive experimental work has detailed how cellular protrusions and morphology vary under different environmental conditions, the mechanistic principles linking protrusive activity to these factors remain poorly understood. Here, we model the extension of actin-based protrusions in chemotaxis as an optimisation problem, wherein cells balance the detection of chemical gradients with the energetic cost of protrusion formation. Our model, built on the assumption of energy minimisation, provides a framework that successfully reproduces experimentally observed patterns of protrusive activity across a range of biological systems and environmental conditions, suggesting that energetic efficiency may underpin the morphology and chemotactic behaviour of motile eukaryotic cells. Additionally, we leverage the model to generate novel predictions regarding cellular responses to other, experimentally untested environmental perturbations, providing testable hypotheses for future experimental work that may be used to validate and refine the model presented here.
\end{abstract}

\newpage

\section{Introduction}

Chemotaxis is the directed migration of cells in response to chemical gradients and it underpins many biological processes, from the immune response to embryonic development \cite{shellard2016chemotaxis, luster2001chemotaxis}. 
In cancer, the ability of malignant cells to detect and traverse gradients of growth factors and chemokines drives their invasion into surrounding tissues \cite{roussos2011chemotaxis, moore2001role, tanaka2005chemokines, liu2004chemotaxis}. 
In each of these processes, cells extend and retract transient actin-based protrusions at their membrane to detect external chemical gradients and also to generate locomotive forces for migration \cite{chodniewicz2004guiding}. 
Despite their importance in chemotaxis, the regulatory mechanisms that govern protrusive dynamics in motile eukaryotic cells remain incompletely understood. 
Thus far, experimental studies have predominantly been concerned with finding correlations between environmental parameters and protrusion formation \cite{yeung2005effects, lo2000cell, ghose2022orientation, wu2012gradient, andrew2007chemotaxis}, rather than elucidating the mechanistic bases of these relationships. 
As such, the effect on protrusive activity of factors such as the magnitude of external chemoattractant gradients and the stiffness of the substrate through which cells migrate remains poorly understood. 
A better understanding of how various environmental factors regulate the protrusive activity of cells would offer insights into how cell motility may be controlled \textit{in vivo}, and hence how, for example, immune responses can be expedited, or metastasis inhibited. 

\subsection{The biology of eukaryotic membrane protrusions}

Cells employ a variety of actin-based membrane protrusions to navigate their surroundings, each distinguished by their morphology and underlying actin architecture \cite{mitchison1996actin}. 
The classification of these actin-based protrusions varies significantly across the literature and is, in general, poorly defined;  
the term \textit{pseudopod} is often used as an umbrella term for various actin-based protrusions. Of the various types of pseudopod characterised in the literature, \textit{lamellipodia} are perhaps the most clearly defined --- broad, sheet-like protrusions composed of densely interwoven actin filaments \cite{condeelis2001lamellipodia}. 
Lamellipodia are typically formed at the leading edge of cells and are most easily observed in two-dimensional migration assays in which cell spreading is more pronounced. 
The primary function of lamellipodia is to form adhesions to the surrounding extracellular matrix (ECM) and to facilitate the movement of cells in the direction of micro-environmental (often chemical) gradients. 
In contrast, \textit{filopodia} are slender protrusions that act primarily as sensory and exploratory structures for cells, allowing for the detection of external signals \cite{wood2002structures}. 
These structures can also be used to form focal adhesions with the ECM that act as anchors to facilitate cell movement. 
For generality and simplicity, in this work we abstract the various categories of pseudopodia referenced in the literature to consider a general class of protrusions extended at the membrane of the cell that are used for both chemical gradient detection and the formation of focal adhesions with the ECM that facilitate movement. 

The formation of many cellular protrusions is driven by actin polymerisation \cite{mogilner1996cell}. 
However, the role of exogenous signals in their formation is a point of controversy in the experimental literature \cite{insall2010understanding}. 
In traditional \textit{chemotactic compass} models, ligand binding at cell surface receptors induces the formation of protrusions by activating intracellular signalling pathways \cite{rickert2000leukocytes}. 
Signalling cascades stimulate the nucleation of actin monomers that subsequently polymerise into filaments bound by fascin \cite{heckman2013filopodia, castellano1999inducible}. As these actin filaments work to deform the inner cell membrane, an outward force is generated that leads to the emergence of a protrusion \cite{ji2008fluctuations}. 
In this regime, cells are polarised in the direction of a chemical gradient, as the formation of protrusions is favoured in regions of high chemical concentration where ligand-receptor binding occurs at a higher rate. 
Conversely, in \textit{pseudopod-centred} models of chemical-induced motility, the initiation of protrusions is independent of external signalling \cite{insall2010understanding}. 
Instead, \textit{de novo} protrusions are formed at random positions along the cell membrane. 
Once formed, exogenous signalling stabilises, grows, and divides pseudopodia preferentially in regions of high chemical concentration to polarise a cell in the direction of a chemical gradient, as in traditional compass-based models. 
Experimental evidence exists to support both of these conflicting frameworks (for a review of studies regarding compass- and pseudopod-centred models, see \cite{insall2010understanding}).

Once stabilised, cellular protrusions form focal adhesions to the ECM, anchoring them in place\cite{stephens2008moving}. 
Following this, actomyosin contractions in the cell generate forces \cite{pandya2017actomyosin} that propel the cell body in the direction of pseudopod-ECM focal adhesion sites. 
Concurrently, the disassembly of rear focal adhesions facilitates the directed migration of the cells. 
Directionality and polarity in the cell, both essential for efficient chemotaxis, are maintained through a combination of signalling pathways and cytoplasmic flow, resulting in a cycle of extension, adhesion, and retraction that facilitates biased motility in the direction of chemical gradients \cite{stock2009chemotaxis}.

\subsection{The study of eukaryotic membrane protrusions}

Extensive experimental work has revealed the complex interplay between the dynamics of actin-based protrusions and environmental parameters. 
Cells modulate their protrusive activity based on multiple factors, including the mechanical properties of the substrate \cite{discher2005tissue, yeung2005effects, stroka2009neutrophils, wang2020effect} and the steepness of the chemical gradient along which they migrate \cite{segall1993polarization, tweedy2013distinct, bosgraaf2008pi3}. 
However, while studies of this kind have helped to elucidate the environmental factors that influence the protrusive activity of cells, it is not clear mechanistically how these factors drive these morphological changes. 

To move beyond associative relationships and probe the mechanistic basis of these phenomena, mathematical modelling offers a valuable complement to experimental methods. Mathematical models of cellular protrusions and motility span scales from single-filament dynamics to whole‐cell behaviour (see extensive reviews by Carlsson et al. \cite{carlsson2008mathematical}, Holmes and Edelstein-Keshet \cite{holmes2012comparison}, and Danuser et al. \cite{danuser2013mathematical}). 
Foundational studies established the Brownian ratchet framework, first formulated by Peskin et al. \cite{peskin1993cellular} and extended by Mogilner et al. \cite{mogilner1996cell}, to explain how actin polymerisation generates membrane forces, while classical one-dimensional models, such as those by Lauffenburger et al. \cite{lauffenburger1989simple} and Dimilla et al. \cite{dimilla1991mathematical}, combine viscoelastic cytoskeletal representations with receptor–ligand kinetics to illustrate how extracellular cues influence protrusive activity. 
More recent three-dimensional and multiscale approaches, including \textit{in silico} models of mesenchymal chemotaxis \cite{ribeiro2017computational, moreno2015fibroblast} and cellular Potts models \cite{maree2006polarization}, integrate molecular-level actin assembly with cell–matrix interactions and chemosensing, reproducing observed cell morphologies and migratory persistence \cite{roberts2000acting}. 
In recent studies, mathematical models of chemotaxis have been used to predict pseudopod-splitting strategies in shallow gradients and optimal morphologies in chemotactic migration \cite{alonso2025persistent, mou2025optimal}. 
However, no general mathematical framework has yet been constructed to study the energy-information trade-off in membrane protrusions across a range of biological contexts and systems. 

Here, we propose a novel theoretical framework that formulates the protrusive dynamics of eukaryotic cells in chemotaxis as an energy optimisation problem. 
The model posits that, during chemotaxis, eukaryotic cells balance the benefits of external chemical gradient detection against the energetic costs of membrane extension and subsequent movement, and hence adopt a protrusive phenotype that yields the highest energetic efficiency. 
Unlike previous modelling efforts, our approach explicitly integrates the energetic costs of protrusion formation with gradient sensing, providing a general predictive framework for protrusive activity across diverse environments.
In what follows, we begin by developing the mathematical foundations of this optimisation model, combining a simple two-dimensional model for chemotaxis with a model for the energy expended upon the extension of protrusions, based on linear elasticity theory. 
We then analyse this model, and the protrusive phenotype that maximises energetic efficiency, to predict the variation in protrusive activity in response to a range of environmental factors. 
Where possible, we compare our predictions with corresponding experiments, and where not, we provide hypotheses that can be tested with future experiments. 

\section{Model}
\label{section:Model derivation}

\begin{figure}[h!]
    \centering
    \includegraphics[width=\linewidth]{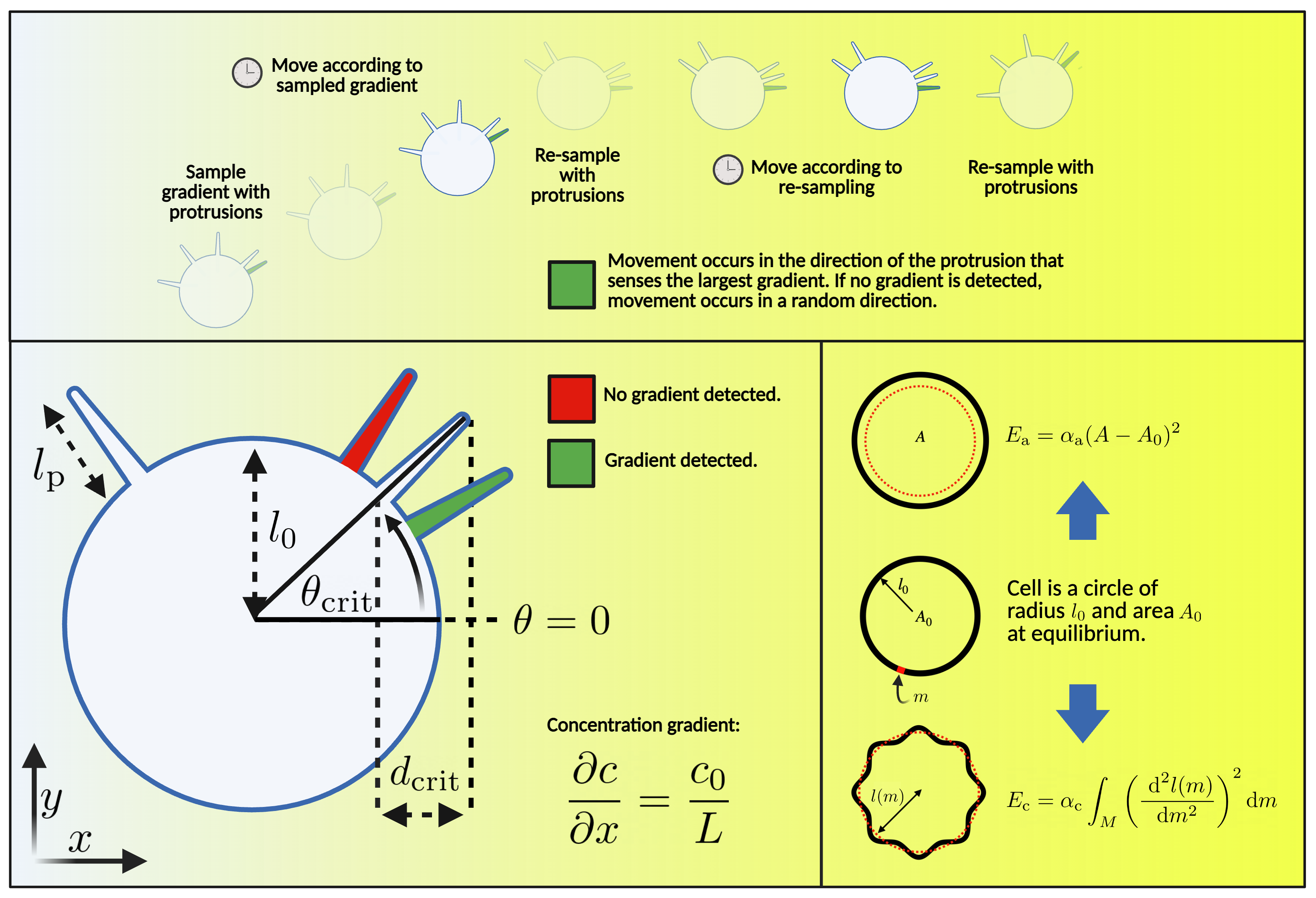}
    \caption{Schematic of the two-dimensional chemotaxis model and the energetic costs of protrusion extension. In the two-dimensional model of chemotaxis, a cell, initially of radius $l_0$, samples its environment with $n_\text{p}$ protrusions at a discrete number of time points (upper panel). Each protrusion is of length $l_\text{p}$, and protrusions are oriented at angles $\{\theta_{i}\}$, sampled independently from an underlying distribution $f(\theta)$ for $\theta \in [0, \pi]$ (neglecting the lower half plane due to symmetry). If the smallest of the angles sampled, $\theta_{\text{min}}$, satisfies $\theta_{\text{min}} \leq \theta_{\text{crit}}(l_\text{p})$, then movement occurs in the direction of the protrusion extended at $\theta_{\text{min}}$. If this condition is not met, then the cell moves in a random direction, such that its expected distance moved along the chemical gradient is zero. The energetic costs of membrane deformation are partitioned into an energy cost due to area change in the cell and an energy cost due to increased membrane curvature, such that all possible perturbations on the circle within the two-dimensional plane are penalised.}
    \label{fig:chemotaxisSchematic}
\end{figure}

We consider a cell of radius $l_{0}$ in the two-dimensional plane and represent its membrane as a closed one-dimensional elastic loop \cite{pontes2013membrane, bernal2007mechanical}. 
We wish to study the energy-motility trade-off in extending actin-rich protrusions at the cell membrane during chemotactic movement. 
In doing so, we begin by using linear elasticity theory to derive expressions for the energetic costs of cell movement and protrusion formation, before constructing a simple two-dimensional model for chemotaxis, in which we relate the number and length of protrusions extended to the probability of chemical gradient detection and subsequent movement (Figure \ref{fig:chemotaxisSchematic}). 
In the following derivation, we assume that cell movement occurs up concentration gradients, though the model is trivially adapted to consider the converse, which is also a well-documented mechanism of migration (e.g., in the migration of neutrophils \cite{tharp2006neutrophil}).

\subsection{Energetic cost of cell movement}
We begin by calculating the energetic cost of movement for a cell. 
For simplicity, we assume that the actin-based protrusions extended by a cell are all of equal length ($l_{\text{p}}$, say) and that they are extended instantaneously, without incurring an increased energetic cost of movement (and hence, that the area of protrusions is small relative to the equilibrium area of the cell). 
We also assume a micro-environment of constant resistance (for a given value of substrate stiffness, $\beta$), such that the energy expenditure per unit distance moved is constant within the domain. 
In this context, energy expenditure due to cell movement upon the extension of protrusions may be quantified as

\begin{equation}
\label{movementCost}
    E_{\text{m}} = \alpha_{\text{m}}\beta d, 
\end{equation}

\noindent where $d$ is the total distance travelled between the extension and retraction of protrusions and $\alpha_{\text{m}}$ is the energy required to move a unit length through a domain in which $\beta=1$. Here, we emphasise that the model derived in this section is a two-dimensional representation of three-dimensional migration \emph{through} a micro-environment of constant resistance. By contrast, for chemotaxis \emph{on} a substrate, where the substrate offers negligible resistance to motion, the energetic cost in Equation \eqref{movementCost} would be effectively zero.
 
\subsection{Energetic cost of membrane bending}
\label{bendingCost}
In calculating the energetic cost of protrusion extension, we adopt a framework similar to that of Ryan et al. \cite{ryan2017cell} and assume that the energetic cost of extending actin-based protrusions can be decomposed into a cost due to the area change of a cell and a cost due to bending of the cell membrane.
At equilibrium, we assume that the outward force from actin polymerisation inside the cell is balanced by a tensile force from the membrane. 
From force balance, we then calculate the energy expended in actin polymerisation as the energy to deform the cell membrane. 

The cell membrane, initially assumed circular with equilibrium radius $l_0$, is parametrised by a coordinate $m \in [0, P]$, where $P = 2\pi l_0$ is the initial circumference of the membrane. 
The coordinate $m$ relates directly to the angular orientation $\theta$ around the cell, measured anticlockwise from a reference direction (chosen as $\theta = 3\pi/2$ at the bottom of the cell). Thus, we have

\begin{equation}
\theta =  \left(\frac{m}{l_0} + \frac{3\pi}{2}\right)\text{ mod } 2\pi. 
\end{equation}

\begin{figure}
    \centering
    \includegraphics[width=0.25\linewidth]{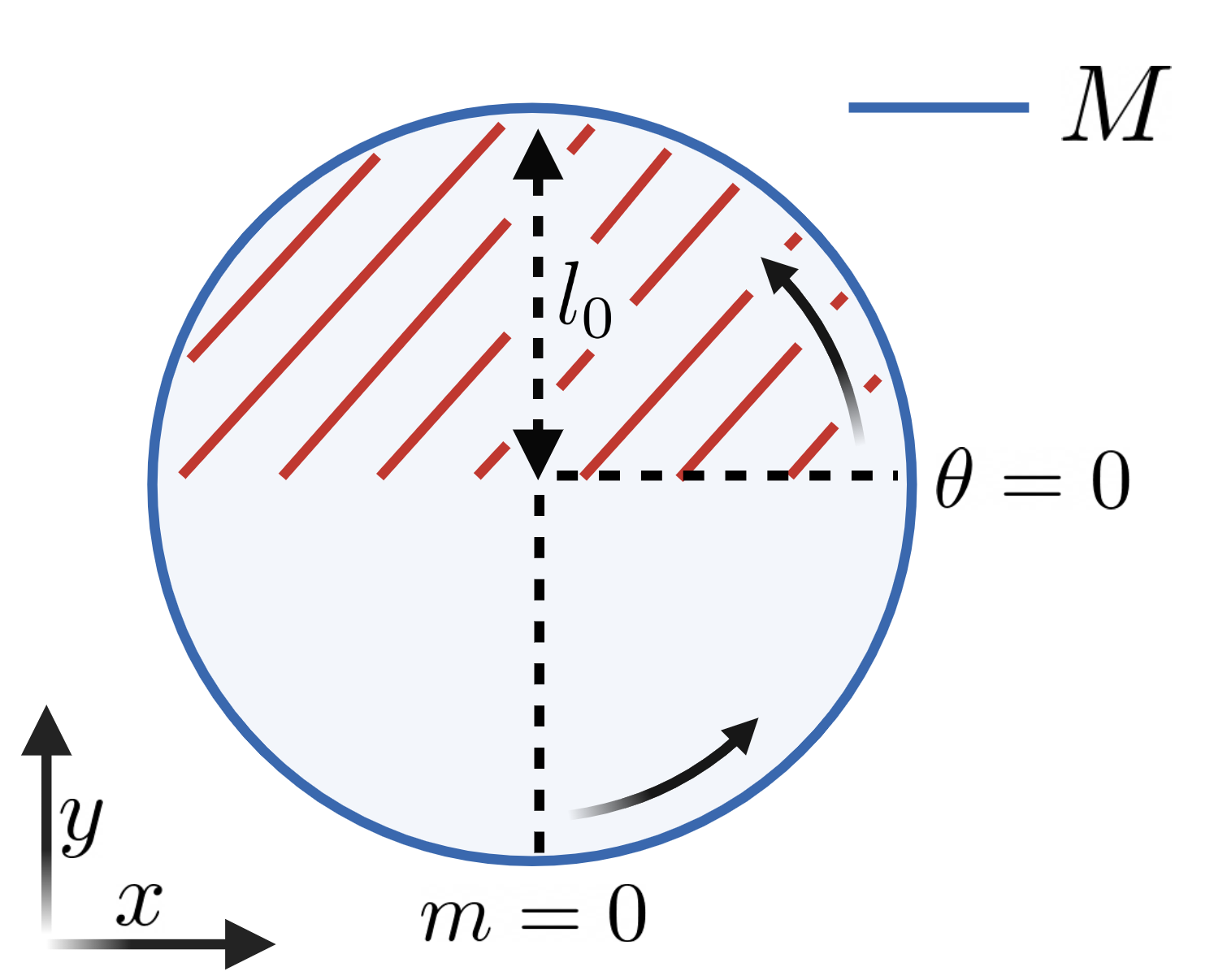}
    \caption{Schematic of the membrane coordinate system. $\theta=0$ denotes the positive $x$-direction and corresponds to the point $m=\pi l_0/2$ on the membrane, $M$. Protrusions are formed in the upper half-plane (dashed red region) to prevent discontinuities at the endpoints of the membrane, $m=0, P$.}
    \label{fig:membraneSchematic}
\end{figure}

To prevent discontinuities at the domain boundaries ($m=0$ and $m=P$), we select this convention so that protrusions are formed away from $m=0$ and $m=P$ (Figure \ref{fig:membraneSchematic}).

When the cell membrane is deformed by extending a protrusion, its curvature, $C$, at a point $m_0$ is given by

\begin{equation}
C(m_0) = \left. \frac{{\rm{d}}^{2}l(m)}{{\rm{d}} m^2}\right\rvert_{m=m_0},
\end{equation}

where $l(m) = l_0 + l_\text{p}(m)$ is the radial distance from the cell centre to point $m$ on the membrane. 
We assume that protrusions are sufficiently small, allowing local segments of the curved membrane to be approximated as linear. 
This simplification justifies taking derivatives directly with respect to the membrane coordinate $m$ without explicitly including higher-order curvature corrections.

Under these assumptions, the potential energy associated with bending the membrane is defined as

\begin{equation}
E_{\text{c}} = \gamma \oint_{M} \left(\frac{\text{d}^{2}l(m)}{\text{d} m^2}\right)^2 \text{d}m,
\end{equation}

where $M$ is the cell membrane and $\gamma$ is an elastic constant corresponding to the membrane bending modulus \cite{miura2020elastic}.

For analytical simplicity, we model protrusions as Gaussian in shape, so that

\begin{equation}
\label{gaussian}
l(m) = l_0 + l_{\text{p}} \exp\left( -\left(\frac{m-m_{\text{c}}}{w_\text{p}}\right)^{2}\right),
\end{equation}

where $l_{\text{p}}$ and $w_\text{p}$ represent the characteristic protrusion length and width, respectively, and $m_\text{c}$ denotes the central location of the protrusion along the membrane.

Calculating the second derivative of $l(m)$ with respect to $m$ indicates that the curvature induced by a protrusion scales linearly with the protrusion length $l_{\text{p}}$. 
Consequently, the bending energy per protrusion scales as $E_{\text{c}} \sim l_{\text{p}}^2$. Furthermore, for well-spaced protrusions of equal length and width, the total bending energy increases linearly with the number of protrusions. 
Combining these scaling relations yields the total energetic cost associated with membrane bending due to $n_\text{p}$ Gaussian protrusions, each of length $l_{\text{p}}$ as

\begin{equation}
\label{curvatureCost}
E_{\text{c}} = \alpha_\text{c} n_\text{p} l_{\text{p}}^{2},
\end{equation}

where $\alpha_{\text{c}}$ represents the energy required to extend a single protrusion of unit length.

\subsection{Energetic cost of cell expansion}

In calculating the energetic cost due to the change in area of a cell upon the extension of actin-based protrusions, we consider an expression that penalises both an increase and decrease in cell area relative to its equilibrium value

\begin{equation}
    E_{\text{a}} = \xi (A - A_0)^2, 
\end{equation}

 where $A$ is the area of a cell after the extension of protrusions, $A_0$ is the baseline area of the cell (its area when no protrusions are extended), and $\xi$ is an elastic constant of expansion and contraction associated with membrane tension. 
 Here, we have assumed the energetic cost to scale quadratically with an increase or decrease in cell area (and hence, cell strain to be sufficiently small that the assumption of linear elasticity holds\cite{postma2004chemotaxis}). 

 When a cell extends a protrusion of a fixed length and width, a constant value is added to the area of the cell ($A_{\epsilon}$, say). 
 We assume that $n_\text{p}$ protrusions of equal size are extended, the corresponding increase in area is given by $n_\text{p}A_{\epsilon}$. The energetic cost resulting from the area change in a cell after the extension of $n_\text{p}$ protrusions of equal size, therefore, scales as $E_{\text{a}} \sim n_\text{p}^{2}$. 
 The area of a Gaussian protrusion scales linearly with its length (for a fixed characteristic width). 
 Consequently, the energetic cost of the area change in a cell also scales quadratically with the length of the protrusions; $E_{\text{a}} \sim l_\text{p}^2$. 

 Combining the scaling relations for $n_\text{p}$ and $l_\text{p}$, the total energetic cost of extending $n_\text{p}$ protrusions, each of length $l_\text{p}$, is

\begin{equation}
    \label{areaCost}
    E_{\text{a}} = \alpha_\text{a} n_\text{p}^2 l_\text{p}^2, 
\end{equation}

where $\alpha_\text{a}$ is the energetic cost due to cell expansion upon extending a single protrusion of unit length. 

\subsection{Total energetic cost of chemical-induced cell motility}
Combining Equations \eqref{movementCost}, \eqref{curvatureCost}, and \eqref{areaCost}, we obtain an expression for the total energy required for a cell to extend $n_\text{p}$ protrusions, each of length $l_\text{p}$, and move a distance $d$ in an ECM of stiffness $\beta$ as 

\begin{equation}
    \label{totalEnergy}
    E = E_{\text{c}} + E_{\text{a}} + E_{\text{m}} = \alpha_\text{c} n_\text{p} l_\text{p}^{2} + \alpha_\text{a} n_\text{p}^2 l_\text{p}^2 + \alpha_{\text{m}}\beta d.  
\end{equation}

\subsection{Expected distance moved in chemical-induced cell motility}
\label{expectedDistance}

We now proceed to formulate a simple two-dimensional representation of eukaryotic chemotaxis to calculate the expected distance moved in chemotaxis as a function of $n_\text{p}$ and $l_\text{p}$ (Figure \ref{fig:chemotaxisSchematic}). 
In defining a chemoattractant landscape, we initially assume that, within the two-dimensional plane, the concentration of the chemical driving movement increases linearly with respect to $x$ and with no dependency on $y$, such that $c(x, y) = c_0x/L$. 
Limits on the accuracy with which cells can detect environmental gradients are well-studied \cite{berg1977physics}. As such, we also assume the existence of a threshold distance in the positive $x$ direction, $d_{\text{crit}}$, which is the minimum distance between the base and the tip of a protrusion that must be spanned in the positive $x$-direction for a chemical gradient to be detected (in Figure \ref{fig:chemotaxisSchematic}, the green protrusion meets this condition, whereas the red protrusion does not).
From the model geometry, $d_{\text{crit}}$, therefore, imposes a minimum concentration difference of $d_{\text{crit}}c_0/L$ between the base and the tip of a protrusion, for a gradient to be detectable. 
The notion of a critical distance, $d_{\text{crit}}$, also affords the definition of a corresponding angle, $\theta_{\text{crit}} \in [0, \pi/2] = \cos^{-1}(l_{\text{p}}/d_{\text{crit}})$, which depends on the length of a protrusion $l_\text{p}$, and determines the maximum angle at which a protrusion can be extended such that a gradient is detected (note here that if $l_\text{p} < d_{\text{crit}}$, no such angle can be defined, meaning that a gradient is never detectable). 
We also assume that protrusions are extended perpendicular to the cell membrane and that the angles at which protrusions are extended are sampled independently from an underlying distribution, $f(\theta), \text{ for } \theta \in [0,\pi]$, where the lower half plane $\theta \in ( \pi, 2\pi )$ is neglected due to up-down symmetry. 
Here, an angle of $\theta = 0$ corresponds to the extension of a protrusion in the positive $x$ direction, while an angle of $\theta = \pi/2$ corresponds to the extension of a protrusion in the positive $y$ direction. 
When a cell extends $n_\text{p}$ protrusions, we assume that all protrusions are the same length and are instantaneously extended in unison (the relaxation of the assumption of equal length across protrusions is the subject of future work within the same framework). 
Furthermore, we assume that a cell moves in the direction of the protrusion that senses the highest environmental concentration gradient (from the geometry of the system, this corresponds to the protrusion extended at the angle closest to $\theta = 0$). 
If a gradient is not detected due to no protrusion spanning a sufficient distance in the positive $x$-direction, then we assume that the cell moves in a random direction. 
We now calculate the expected distance moved by a cell upon the extension of $n_\text{p}$ protrusions of length $l_\text{p}$. 

To calculate the probability of gradient detection in cell movement (the probability that at least one protrusion is extended at an angle lower than $\theta_{\text{crit}}$ in Figure \ref{fig:chemotaxisSchematic}), we derive the probability density function of the minimum angle of $n_\text{p}$ samples from the distribution $f(\theta)$. 
For $n_\text{p}$ independent samples from this distribution, the probability that all angles are greater than a value $\theta$ is given by

\begin{equation}
    \left (1 - \int_{0}^{\theta}f(\theta^{'}) \text{d} \theta^{'} \right)^{n_\text{p}}.
\end{equation}

\noindent From this, the cumulative distribution function of the minimum of these $n_\text{p}$ samples is given by

\begin{equation}
    \label{CDF}
    P(\theta) = 1 - \left (1 - \int_{0}^{\theta}f(\theta^{'}) \text{d} \theta^{'} \right)^{n_\text{p}}.
\end{equation}

\noindent Differentiating Equation \eqref{CDF} with respect to $\theta$, the probability density function of the minimum angle sampled is

\begin{equation}
    p(\theta) = n_\text{p} f(\theta) \left(1 - \int_{0}^{\theta}f(\theta^{'}) \text{d}\theta^{'}\right)^{n_\text{p}-1}.
\end{equation}

\noindent From this expression, we obtain the probability of gradient detection upon the extension of $n_\text{p}$ protrusions of length $l_\text{p}$ as 

\begin{equation}
    \int_{0}^{\theta_{\text{crit}}} p(\theta^{''})\text{d}\theta^{''} = n_\text{p}\bigintsss_{0}^{\theta_{\text{crit}}} f(\theta^{''}) \left(1 - \int_{0}^{\theta^{''}}f(\theta^{'}) \text{d}\theta^{'}\right)^{n_\text{p}-1}\text{d}\theta^{''},
\end{equation} 

\noindent where $\theta_{\text{crit}}$ is the maximum angle for which a protrusion of length $l_{\text{p}}$ can be extended such that a gradient is detected:

\begin{equation}
    \theta_{\text{crit}} = \cos^{-1}\left(\frac{d_{\text{crit}}}{l_\text{p}}\right).
\end{equation}
\\ \noindent We now calculate the expected distance travelled by a cell upon the extension of $n_\text{p}$ protrusions of length $l_\text{p}$. 
If the smallest of the $n_\text{p}$ angles sampled, $\theta_{\text{min}}$, satisfies $\theta_{\text{min}} \leq \theta_{\text{crit}}$, the distance moved by a cell in the positive $x$ direction (the direction of the environmental gradient) is given by

\begin{equation}
    d\cos(\theta_{\text{min}}), 
\end{equation}

\noindent where $d$ is the distance moved by the cell upon the extension of protrusions. 
If the minimum orientation of $n_\text{p}$ protrusions does not meet the threshold for gradient detection ($\theta_{\text{min}} > \theta_{\text{crit}}$), movement is unbiased, and hence, vanishes under expectation. 
Consequently, the expected distance travelled by a cell in the direction of an environmental gradient upon extending $n_\text{p}$ protrusions, each of length $l_\text{p}$, is obtained as 

\begin{equation}
    \label{expectedMovement}
    dn_\text{p}\bigintsss_{0}^{\theta_{\text{crit}}} f(\theta^{''})\cos(\theta^{''}) \left(1 - \int_{0}^{\theta^{''}}f(\theta^{'})\text{d}\theta^{'}\right)^{n_\text{p}-1}\text{d}\theta^{''},
\end{equation}

\noindent where we have conditioned on gradient-induced movement by integrating only from $0$ to $\theta_{\text{crit}}$. 

A final consideration we make in the expression for distance moved is the effect of substrate stiffness and organisation on the speed at which a cell can move (and hence, the distance moved between the extension and retraction of protrusions). 
In the model, we characterise the ECM in terms of its stiffness ($\beta$), and a parameter that we refer to as topographical permissibility ($\phi$), which determines how permissive to cell migration the ECM is for a given value of $\beta$ (and hence, is a parameter that captures properties such as the degree to which fibres in the ECM are aligned). We then define a function, $g(\beta, \phi)$, that modulates the distance moved by a cell between the extension and retraction of protrusions, and depends on the stiffness ($\beta$) and topography ($\phi$) of the surrounding ECM. 
As such, Equation \eqref{expectedMovement} can be modified to account for substrate stiffness and permissibility as

\begin{equation}
    \label{expectedMovementSubstrate}
    g(\beta, \phi)dn_\text{p}\bigintsss_{0}^{\theta_{\text{crit}}} f(\theta^{''})\cos(\theta^{''}) \left(1 - \int_{0}^{\theta^{''}}f(\theta^{'})\text{d}\theta^{'}\right)^{n_\text{p}-1}\text{d}\theta^{''}.
\end{equation}

At this point, we do not prescribe a functional form for $g(\beta, \phi)$, though in subsequent sections, we make phenomenological selections for $g(\beta, \phi)$ to study how substrate stiffness and topography affect protrusive activity in a range of biological scenarios. 

\subsection{Final expression for the expected distance travelled per unit energy expended in environmental sensing and cell movement}
Combining Equations \eqref{totalEnergy} and \eqref{expectedMovementSubstrate}, we arrive at an expression for the expected distance, $d_x$, travelled up a fixed chemical gradient in chemotaxis per unit energy expended by a cell ($E$) upon the extension of $n_\text{p}$ protrusions, each of length $l_\text{p}$ as

\begin{equation}
    \label{DTERDimensional}
    \frac{\mathbb{E}(d_x)}{E} = \frac{g(\beta, \phi)dn_\text{p}\displaystyle\int_{0}^{\theta_{\text{crit}}} f(\theta^{''})\cos(\theta^{''}) \left(1 - \int_{0}^{\theta^{''}}f(\theta^{'})\text{d}\theta^{'}\right)^{n_\text{p}-1}\text{d}\theta^{''}}{\alpha_\text{c} n_\text{p} l_\text{p}^{2} + \alpha_\text{a} n_\text{p}^2 l_\text{p}^2 + \alpha_{\text{m}}\beta g(\beta, \phi)d}. 
\end{equation}

Without loss of generality, Equation \eqref{DTERDimensional} may be simplified by fixing $\alpha_\text{c}=1$ and rescaling all other terms in Equation \eqref{DTERDimensional}, such that

\begin{equation}
    \label{DTER}
    \frac{\mathbb{E}(d_x)}{E} = \frac{g(\beta, \phi)dn_\text{p}\displaystyle\int_{0}^{\theta_{\text{crit}}} f(\theta^{''})\cos(\theta^{''}) \left(1 - \int_{0}^{\theta^{''}}f(\theta^{'})\text{d}\theta^{'}\right)^{n_\text{p}-1}\text{d}\theta^{''}}{n_\text{p} l_\text{p}^{2} + \alpha_\text{a} n_\text{p}^2 l_\text{p}^2 + 
    \alpha_{\text{m}} \beta g(\beta, \phi) d}. 
\end{equation}

We henceforth refer to Equation \eqref{DTER} as the distance-to-energy ratio (DTER) of cell movement.

\section{Results}
\label{results}
We now analyse the model as defined in Equation \eqref{DTER} to study the variation in the optimum protrusive phenotype (the combination of $n_{\text{p}}$ and $l_{\text{p}}$ that maximise the DTER), as a function of various environmental parameters and cellular behaviours. 

\subsection{Shallow chemical gradients drive a transition from bet-hedging to speculative phenotypes}
\label{shallow}
The first parameter we consider in the study of Equation \eqref{DTER} is the threshold distance for the detection of a chemical gradient by the cell, $d_{\text{crit}}$. 
This parameter can be thought of as representing the steepness of the chemical gradient in the domain, with higher threshold distances representing a shallower chemical gradient as protrusions must span a larger distance for a gradient to be detected. 

In analysing the effect of $d_{\text{crit}}$ on the optimal protrusive phenotype, we assume that $f(\theta)$, the probability distribution that governs the orientation of the protrusions extended at the cell membrane, is the uniform distribution $\mathrm{U}[0, \pi]$, such that

\begin{equation}
    f(\theta) = \frac{1}{\pi}. 
\end{equation}

This form of $f(\theta)$ represents a pseudopod-centred view of chemotaxis, where the generation of actin-driven protrusions is independent of the environmental chemical landscape within which a cell lies. 
However, we will later consider biases in this probability distribution, to compare the pseudopod-centred model with traditional chemotactic compass models (Section \ref{angleDistribution}). 

In selecting $g(\beta, \phi)$, we again choose a phenomenological expression that penalises movement in substrates of extreme stiffness, and for which the topographical permissibility, $\phi$, linearly increases the distance moved by a cell per unit of time. 
In very soft substrates, cells cannot generate sufficient traction forces for movement \cite{lo2000cell}. 
In contrast, on substrates of very high stiffness, cell motility is impeded by the physical barrier of the dense substrate \cite{lange2013cell}. 
A functional form of $g(\beta, \phi)$ that satisfies these constraints is

\begin{equation}
    \label{g(beta, phi)}
    g(\beta, \phi) = \frac{\phi\beta}{k}\exp\left(-\left(\frac{\beta}{k}\right)^{2}\right),
\end{equation}

which peaks at an intermediate substrate stiffness of $\beta = k/\sqrt{2}$. With these two assumptions, Equation \eqref{DTER} becomes

\begin{equation}
    \label{DTER_substituted}
    \frac{\mathbb{E}(d_x)}{E} = \frac{\phi \beta d n_\text{p}}{k\pi} \frac{ \exp\left( -\left( \frac{\beta}{k} \right)^2 \right)\displaystyle\int_{0}^{\theta_{\text{crit}}} \cos(\theta'') \left( 1 - \frac{\theta''}{\pi} \right)^{n_\text{p} -1} \, \rm{d}\theta'' }{ n_\text{p} l_\text{p}^2 + \alpha_a n_\text{p}^2 l_\text{p}^2 + \alpha_{\text{m}} \beta \left( \frac{\phi \beta}{k} \exp\left( -\left( \frac{\beta}{k} \right)^2 \right) \right) d}.
\end{equation}

In the absence of experimental parameterisation, we set the parameters $\alpha_{\text{a}}$,  $\alpha_{\text{m}}$, and $\phi$ to unity and select values of $\beta=10^2$ and $k=10^2/\sqrt{2}$ that constrain the optimum values of $n_{\text{p}}$ and $l_{\text{p}}$ to lie between 1 and 10 for the parameter ranges considered. 
However, in Appendix A, we discuss the results of a general model parameter sweep to study the effect of each parameter choice on the optimal values of $n_\text{p}$ and $l_\text{p}$.

\begin{figure}
    \centering
    \includegraphics[width=\linewidth]{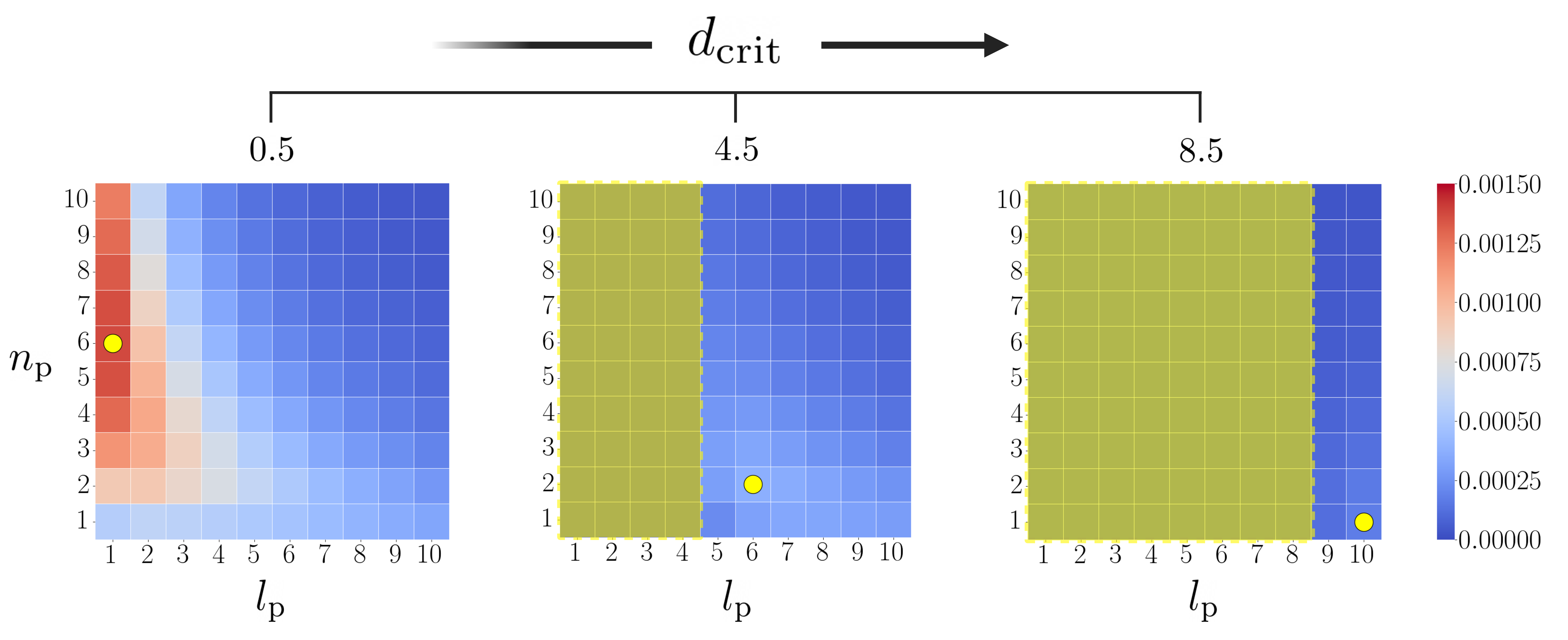}
    \caption{Heatmaps showing the evolution of the DTER with respect to $d_{\text{crit}}$ (the steepness of the chemical gradient within the domain). When $d_{\text{crit}}=0.5$, the chemical gradient within the domain is of a large magnitude, such that the optimal strategy (yellow dot) is to sample multiple directions with short protrusions. Shaded yellow regions show parameter values in which gradient detection is not possible ($d_{\text{crit}} > l_{\text{p}}$). As $d_{\text{crit}}$ is increased towards $d_{\text{crit}}=8.5$, representing increasingly shallower chemical gradients, the optimal strategy shifts to the extension of a lower number of protrusions, each of a longer length. All figures show numerical solutions of Equation \eqref{DTER} with $\alpha_{\text{m}} = \alpha_{\text{a}} = \phi = 1$, $\beta=10^2$, and $k = 10^2/\sqrt{2}$.}
    \label{fig:gradientMagnitude}
\end{figure}

Varying the threshold distance protrusions must span in the direction of the environmental gradient reveals a transition between phenotypes as chemical gradients become increasingly shallow (Figure \ref{fig:gradientMagnitude}). 
For low values of $d_{\text{crit}}$ (i.e.~steep chemoattractant gradients), the optimum protrusive phenotype is one in which cells extend multiple short protrusions. 
Only short protrusions are required due to easily detectable chemical gradients, and multiple protrusions are extended to ensure that the direction of movement is optimised with respect to chemical gradients. 
In contrast, in shallow gradients, where long protrusions must be extended in order to detect a chemical gradient, the optimum number of protrusions to extend is $n_{\text{p}} = 1$, due to the energetic costs of extending increasingly longer protrusions.
This result highlights two distinct strategies for gradient detection, \textit{bet-hedging} phenotypes, wherein risk is spread across multiple directions and \textit{speculation} in shallow gradients, wherein all risk is placed in sampling a single direction. 

\textit{In vivo} experimental evidence for the transition between speculation and bet hedging (Figure \ref{fig:gradientMagnitude}) lies in the cranial neural crest, wherein streams of cells migrate collectively, guided by external gradients in vascular endothelial growth factor (VEGF) \cite{mclennan2015vegf}. 
\textit{In vivo}, the morphology of cranial neural crest cells varies along streams \cite{teddy2004vivo}. 
At the leading edge of streams, where gradients in VEGF are steep, cranial neural crest cells adopt \textquotesingle{}hairy\textquotesingle{} phenotypes, with multiple protrusions extended in many different directions. 
At the back of streams, where gradients in VEGF are extremely shallow, cells instead adopt \textquotesingle{}bipolar\textquotesingle{} phenotypes, becoming highly polarised and extending few, long protrusions. 
Furthermore, the transition between the bet-hedging and speculative phenotypes predicted here has also been observed \textit{in vitro} for yeast cells (\textit{Saccharomyces cerevisiae}) exposed to spatial gradients of a mating factor \cite{segall1993polarization}, where a loss of polarisation is associated with exposure to steeper chemical gradients.

However, in other contexts, the converse is true and cells in steep chemical gradients develop a single dominant leading edge and robust polarisation, while shallow gradients yield more exploratory, branched pseudopodial activity (e.g.~in endothelial cells exposed to gradients in VEGF \cite{shamloo2008endothelial}). 
We propose that this discrepancy can be reconciled by considering the distance travelled by neural crest cells relative to other migratory cell populations. 
\textit{In vivo}, cranial neural crest cells must travel distances upwards of $1000\mu $m to colonise the branchial arches and facilitate proper facial patterning \cite{mclennan2010vascular}. 
This distance is significantly larger than many other cell types, and is the largest distance for any embryonic cell type in the vertebrate embryo \cite{broders2016control}. 
As such, energetic constraints in gradient detection may be more of a consideration for neural crest cells than the other cell types studied in this context, for which the rapid detection of gradients and subsequent migration may be prioritised (e.g. in wound healing). 
 
\subsection{Cell spreading exhibits a biphasic relationship with substrate stiffness \textit{in vivo}}
\label{substrateStiffness}

In Equation (\ref{DTER_substituted}), the stiffness of the substrate through which migration occurs is represented by the parameter $\beta$. The parameter $\phi$ captures all properties of the substrate that modify a cell's capacity for motility but do not affect the energetic cost of movement through the domain (e.g.~the degree of fibre alignment in the ECM). Within this framework, we assume that a denser or stiffer micro-environment incurs an increased energetic cost of movement in cells. We also assume that substrate stiffness modulates the distance moved in chemical-induced motility due to a variation in the traction force a cell can generate and the mechanical impedance of cell movement in stiff substrates. Here, we fix $\beta$ to be constant within the domain, and hence, assume the stiffness of the substrate to be independent of the movement stimulus concentration profile.

We now study the effect of substrate topography on cell morphology by studying the combination of protrusion lengths and numbers that maximise the DTER for a range of substrate stiffnesses, $\beta$, and substrate permissibilities, $\phi$. As in Section \ref{shallow}, we assume that the distribution governing the orientation of protrusions is the uniform distribution, $\mathrm{U}[0, \pi]$, and that the the parameters $\beta$ and $\phi$ influence motility according to Equation \eqref{g(beta, phi)}. In the absence of experimentally-informed model parameterisation, we fix the membrane-related energetic parameters $\alpha_\text{c}$ and $\alpha_\text{a}$ at unity, and systematically vary $\beta$ and $\phi$, representing a progressive stiffening and aligning of the ECM. 

\begin{figure}
    \centering
    \includegraphics[width = \textwidth]{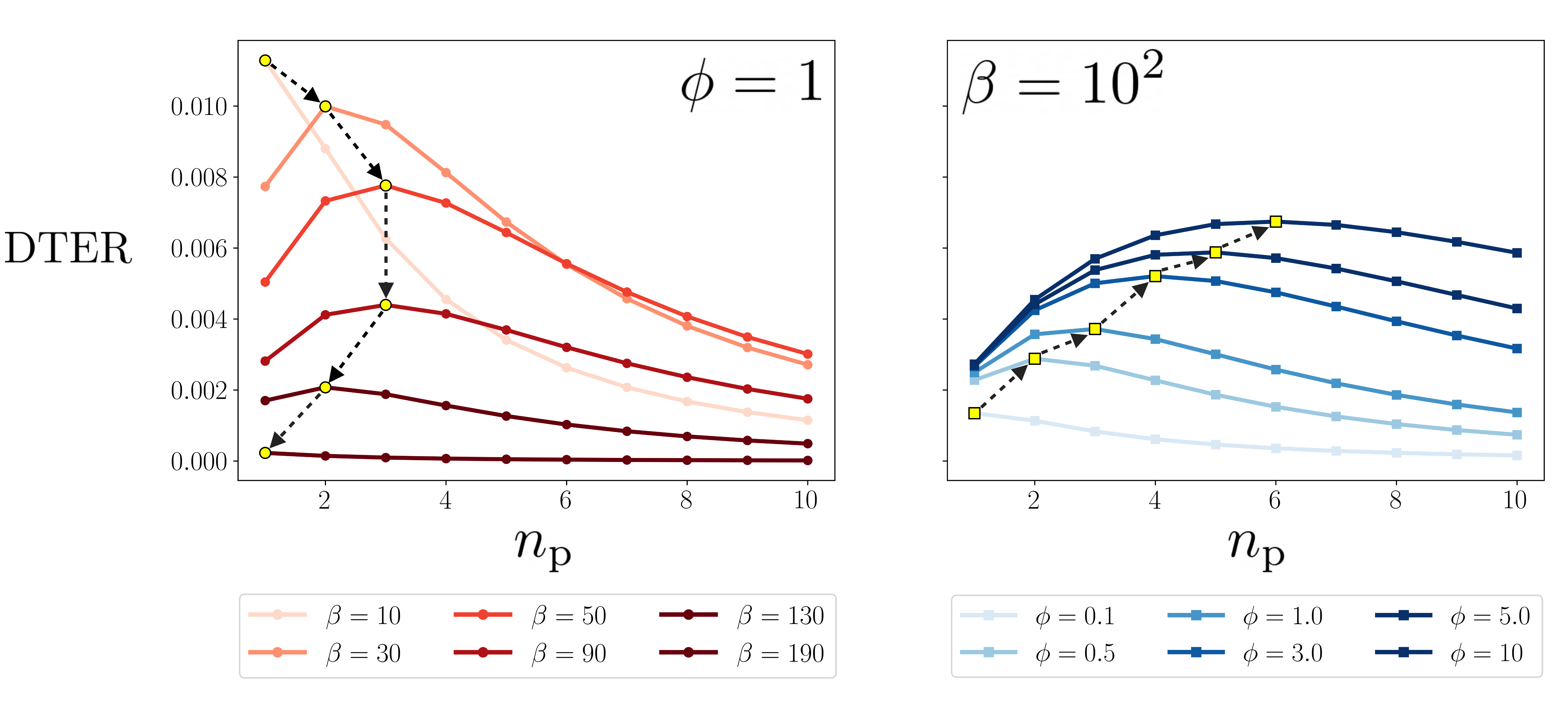}
    \caption{DTER as a function of $n_\text{p}$ for a range of values of $\beta$ and $\phi$ ($\alpha_\text{a} = \alpha_\text{m} = 1, d_{\text{crit}}=0.5, l_\text{p}=1, k=10^2/\sqrt{2}$). For low substrate stiffnesses (small values of $\beta$), model predictions suggest that the optimum number of protrusions to extend is $n_\text{p} = 1$. As substrate stiffness, $\beta$, is increased, the DTER is optimised for larger values of $n_\text{p}$. For very large substrate stiffnesses, the optimum value of $n_\text{p}$ peaks, and subsequently decreases back to a value of $n_\text{p} = 1$. Increasing $\phi$ increases the optimum value of $n_\text{p}$ monotonically. Yellow markers indicate the value of $n_\text{p}^{\text{opt}}$ and arrows connect successive values of $n_\text{p}^{\text{opt}}$ as $\beta$ and $\phi$ are increased.}
    \label{fig:movementCostnp}
\end{figure}

Figure \ref{fig:movementCostnp} shows that protrusive activity is maximised for intermediate values of $\beta$, where the distance migrated in-between sampling times is maximised. Furthermore, increasing $\phi$ increases the optimal value of $n_\text{p}$ monotonically. Taken together, this suggests that optimal protrusive activity in migration is linked to the motile capacity of cells, as determined by the substrate stiffness and topography. This, in turn, suggests that for migration through substrates \textit{in vivo}, cells encountering highly aligned fibre networks and optimal stiffness can harness increased motility to extend numerous protrusions, thereby sampling their surroundings extensively and maintaining directional persistence along environmental concentration gradients. Conversely, in regions where the ECM is either too compliant or overly rigid, such as loosely organised interstitial spaces or densely cross-linked fibrotic tissue, cells may reduce protrusive sampling, favouring a minimal number of projections to conserve energy. These results reconcile conflicting trends observed \textit{in vitro}, where in some cases, a stiffening of three-dimensional substrates causes enhanced cell spreading\cite{mason2013tuning}, whereas in other studies, the converse is true \cite{stowers2015dynamic}. The findings presented here suggest that these experiments may be sampling distinct regions of the biphasic dashed $\beta$-varying trajectory in Figure \ref{fig:movementCostnp}, and hence, observing opposing relationships between substrate stiffness and cell spreading. In both cases, if a wider range of stiffnesses were considered, our analysis predicts that the biphasic relationship predicted here would be observed. This hypothesis is also supported by additional experiments in which it has been observed that protrusion rates in motile cells peak at intermediate collagen densities \cite{fraley2015three}. 

The results presented in this section, therefore, suggest that substrate topography is a major determinant of protrusive activity in motile cells, and that this relationship may be understood through the trade-off in the energy expended in environmental sampling and the distance moved upon sampling. In overly stiff or compliant substrates, the distance moved through the substrate upon sampling is minimal, such that additional protrusions confer little motility benefit but incur a high energetic cost. Conversely, when the matrix composition is of an intermediate stiffness, allowing for longer migratory distances between sampling, cells can afford to generate, and indeed benefit from extensive exploratory protrusions. If a cell is poised to travel a greater distance, it is energetically favourable to optimise the direction of movement by deploying numerous projections to accurately map out the local chemoattractant profile, thus increasing the likelihood of movement in the direction of the external chemical gradient.

The stiffness‐dependent protrusion dynamics predicted by the model also raise intriguing questions about how cancer cells exploit invadopodia \cite{weaver2008invadopodia} and matrix metalloproteinases (MMPs) \cite{johnson1998matrix} to tune their own effective substrate mechanics. Invadopodia are actin‐rich, protease‐laden projections that locally degrade the ECM via MMP secretion. A coupling between protrusion density along the cell membrane and ECM degradation suggests the existence of a feedback loop between protrusive activity and substrate stiffness, wherein malignant cells extend invadopodia to degrade overly stiff substrates, and hence increase their motile capacity, which in turn, changes the optimal protrusive activity predicted by the model considered here. When stiffness is locally decreased to lie within the mechanical window that optimises motile capacity, cells may then downregulate invadopodia formation to conserve resources while maintaining exploratory protrusions for chemoattractant sampling. In contrast, when the ECM falls below the lower stiffness threshold,  proteolysis would compromise traction forces and impede movement by further softening the substrate, such that cells may suppress invadopodia, yet continue to deploy non-proteolytic protrusions to survey their surroundings. Embedding this antagonistic regulation, where $\beta$ and $\phi$ dynamically evolve in response to MMP‐mediated remodelling and protrusive sampling, into the DTER framework may capture how cancer cells iteratively sculpt their micro‐environment, balancing ECM proteolysis and motility energetics to carve permissive invasion tracks, sustain directional persistence, and maximise invasive potential in mechanically heterogeneous tissues. 

In addition to modulating local stiffness, MMP activity mobilises matrix-sequestered growth factors and chemoattractants \cite{page2007matrix}, thereby establishing steep concentration gradients that can bias protrusion density and orientation. Such biochemical release mechanisms introduce an additional layer of protrusive regulation, operating in parallel with stiffness-dependent feedbacks to shape migratory dynamics. Invadopodia, therefore, function as structures that modify both the mechanical compliance of the ECM and the distribution of guidance cues in the surrounding micro-environment. Extending the DTER framework to include these coupled mechano-chemical interactions would enable analysis of how cancer cells synchronise proteolytic remodelling, mechanical adaptation, and chemotactic sampling to optimise invasion strategies.

\subsection{Integration of signals between actin-based protrusions is optimal in well-sampled environments}
\label{section:ProjectionThresholding}

In many eukaryotic cell types, there is evidence for signal integration across the cell membrane \cite{postma2004chemotaxis}, such that chemotactic steering may be achieved not only by considering gradients detected by individual protrusions, but also by the integration of signals across the cell membrane to compare chemoattractant concentrations sampled at different receptor sites \cite{devreotes2003eukaryotic, devreotes2017excitable, swaney2010eukaryotic}. In the context of the model considered here, this suggests an alternative model of environmental sampling, wherein the detection of a chemical gradient depends not on the concentration difference measured by an individual protrusion, but rather on the \textit{total} concentration difference spanned by a cell upon the extension of multiple protrusions. In this section, we formalise this notion in terms of the model by defining
\[
\theta_{\min} \;=\;\min_{1\le i\le n_{\rm{p}}}\theta_i,
\qquad
\theta_{\max} \;=\;\max_{1\le i\le n_{\rm{p}}}\theta_i.
\]
The tip of each protrusion of length \(l_\text{p}\) measures the chemoattractant concentration at a distance $(l_0 + l_\text{p})\cos(\theta_i)$ in the direction of the gradient relative to the centre of the cell, where $\theta_{i}$ is the orientation of protrusion $i$. As such, the maximum distance spanned in the $x$-direction by the tips of any two protrusions is given by

\begin{equation}
    \label{d_int}
    d_{\text{int}}=(l_0 + l_p)\cos(\theta_{\min})-(l_0 + l_p)\cos(\theta_{\max}).
\end{equation}

As before, we assume that upon the extension of $n_{\text{p}}$ protrusions at the cell membrane, the orientations of the protrusions are determined by independent samples from the angular distribution $f(\theta)$;
\[
\{\theta_i\}_{i=1}^{n_\text{p}} \;\overset{\mathrm{i.i.d.}}{\sim}\; f(\theta), 
\quad \theta\in[0,\pi].
\]

Writing the cumulative distribution function for $f(\theta)$ as
\begin{equation}
    F(\theta)=\int_0^\theta f(\phi)\,\mathrm{d}\phi,
\end{equation}
the joint probability density function of the smallest and largest orientations may be expressed as
\begin{equation}
p_2(\theta_{\min},\theta_{\max})
= n_{\text{p}}(n_{\text{p}}-1)\,\bigl[F(\theta_{\max})-F(\theta_{\min})\bigr]^{\,n_{\text{p}}-2}\,
f(\theta_{\min})\,f(\theta_{\max}),
\quad 0\le\theta_{\min}<\theta_{\max}\le\pi,
\end{equation}

\noindent where $n_{\mathrm p}(n_{\mathrm p}-1)$ counts the choices of which two samples attain the minimum and maximum, $f(\theta_{\min})$ and $f(\theta_{\max})$ are the endpoint densities, and $\bigl[F(\theta_{\max})-F(\theta_{\min})\bigr]^{\,n_{\mathrm p}-2}$ is the probability that the remaining $n_{\mathrm p}-2$ samples lie within $(\theta_{\min},\theta_{\max})$.

\begin{figure}
    \centering
    \includegraphics[width=\linewidth]{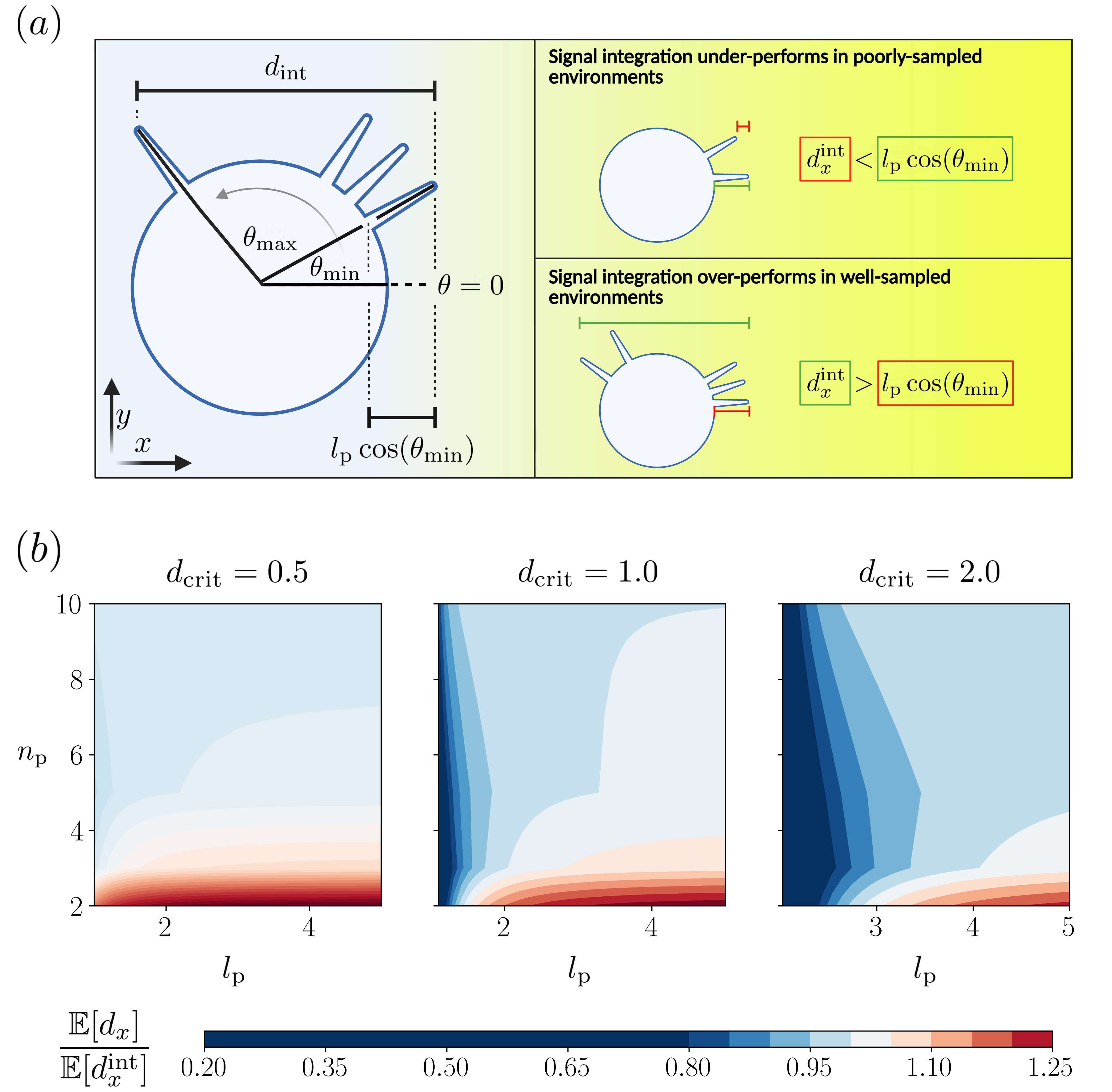}
    \caption{\textbf{(a)} Schematic of the original sensing mechanism and the protrusion-integrated sensing mechanism. In the original mechanism, a chemical gradient is detected if the protrusion extended at the lowest angle ($\theta_{\rm{min}})$ spans a sufficient distance in the positive $x$-direction. In the integrated sensing mechanism, a gradient is detected if the tips of the protrusions extended at the smallest and largest angles ($\theta_{\rm{min}}$ and $\theta_{\rm{max}}$) span a sufficient distance in the $x$-direction ($d_{\text{int}} \geq d_{\text{crit}}$). Model analysis shows that when the local environment is well-sampled (i.e. $n_{\text{p}}$ is sufficiently high), the integrated sensing mechanism outperforms the single-protrusion sensing mechanism. Conversely, when the environment is poorly-sampled, the original single-protrusion detection mechanism outperforms signal integration for sufficiently large values of $l_{\text{p}}$. \textbf{(b)} Heatmaps of the ratio of the expected distances moved under the integrated versus single-protrusion sensing mechanisms, computed for three detection thresholds ($d_{\mathrm{crit}} = 0.5, 1.0, 2.0$). For all panels, the variables not swept over are fixed at $\alpha_{\mathrm{m}}=\alpha_{\mathrm{a}}=\phi=1$, $\beta=10^2$, $k=10^2/\sqrt{2}$, and $l_{0}=10$.} 
    \label{fig:signalIntegration}
\end{figure}

Now, rather than requiring the $x$-component of one protrusion to exceed a fixed detection length, we instead impose a thresholding condition by requiring the difference in these projections to exceed a threshold distance, \(d_{\rm crit}\),
\begin{equation}
    d_{\text{int}}\ge d_{\rm crit}, 
\end{equation}

where $d_{\text{int}}$ is the maximum distance in the $x$-direction spanned by the tips of any two protrusions, as defined in Equation \eqref{d_int} and visualised in Figure \ref{fig:signalIntegration}(a). 

If this condition holds, the cell moves a distance \(d\) in the direction of the leading protrusion (\(\theta_{\min}\)); otherwise its expected drift vanishes (Figure \ref{fig:signalIntegration}(a)).  From this, the expected drift in the $x$-direction upon the extension of protrusions is given by
\begin{equation}
\label{eq:ExpectedDriftProj}
\mathbb{E}[d_x^{\text{int}}]
= dg(\beta, \phi)
\int_{0}^{\pi}\!\!
\int_{\theta_{\min}}^{\pi}
\cos(\theta_{\min})
\;\mathds{1}_{\{\,(l_0 + l_\text{p})(\cos(\theta_{\min})-\cos(\theta_{\max}))\ge d_{\rm crit}\}}
\;p_2(\theta_{\min},\theta_{\max})
\;\mathrm{d}\theta_{\max}\,\mathrm{d}\theta_{\min}.
\end{equation}

In comparing this sampling strategy with the original sampling strategy defined in Section \ref{section:Model derivation}, we consider the fraction
\begin{equation}
\frac{\mathbb{E}[d_x]}{\mathbb{E}[d_x^{\text{int}}]} = \frac{n_\text{p}\displaystyle\int_{0}^{\theta_{\text{crit}}} f(\theta^{''})\cos(\theta^{''}) \left(1 - \int_{0}^{\theta^{''}}f(\theta^{'})\text{d}\theta^{'}\right)^{n_\text{p}-1}\text{d}\theta^{''}}{
\displaystyle\int_{0}^{\pi}\!\!
\displaystyle\int_{\theta_{\text{min}}}^{\pi}
\cos(\theta_{\min})
\;\mathds{1}_{\{\,(l_0 + l_\text{p})(\cos(\theta_{\min})-\cos(\theta_{\max}))\ge d_{\rm crit}\}}
\;p_2(\theta_{\min},\theta_{\max})
\;\mathrm{d}\theta_{\max}\,\mathrm{d}\theta_{\min}}, 
\end{equation}

which computes the ratio of expected distances moved in the positive $x$-direction in both of the sampling mechanisms considered here for a fixed $n_{\text{p}}$ and $l_{\text{p}}$. In order to compare the relative efficacy of these two mechanisms, we perform a parameter sweep over $n_\text{p}$, $l_\text{p}$, and $d_{\text{crit}}$ to determine the contexts in which either of these two mechanisms are optimal. In the proceeding analysis, we once again assume that the angular distribution, $f(\theta)$, is the uniform distribution $\mathrm{U}[0, \pi]$. 

The model reveals a consistent pattern across a range of chemoattractant threshold distances. When cells extend only a few long protrusions, the original single-protrusion measurement mechanism outperforms the signal integration mechanism in terms of expected forward drift per unit of energy expended (Figure \ref{fig:signalIntegration}(b)). In this regime, there is a moderate probability of any given protrusion spanning the threshold distance in the positive $x$-direction. However, even if multiple protrusions individually meet this thresholding condition, it may still be the case that the distance between the tips of any two protrusions do not. Furthermore, even if all protrusions are extended in the negative $x$-direction, a positive concentration difference may be detected between the tips of two protrusions, leading to movement down chemical concentration gradients. By contrast, when cells deploy many shorter protrusions, it is unlikely that a single protrusion individually meets the thresholding condition, but it is very likely, that, of the many protrusions extended, the tips of the protrusions extended at the smallest and largest angles span a distance greater than $d_{\text{crit}}$, and that the smallest angle (which we assume determines the direction of movement) points in the approximate direction of the concentration gradient.

Placing these findings in a biological context suggests that cell-type--specific protrusive strategies may reflect adaptations to distinct migratory challenges.  Cells such as neural crest or invasive carcinoma that probe dense matrices with a small number of long filopodia likely benefit from a single‐protrusion detection mechanism, enabling them to migrate towards distant targets once an individual protrusion measures a chemoattractive gradient. In contrast, fibroblasts migrating in complex micro-environments extend numerous pseudopodia and lamellipodia that are distributed broadly over the leading edge. In such a regime, gradient sensing cannot rely on any single protrusion; instead, the cell may integrate signals across many simultaneously extended tips. In this context, signal integration may help to suppress stochastic fluctuations measured by individual protrusions and stabilise migration in shallow or noisy chemoattractant fields.

\subsection{Angular bias in protrusion orientation is optimal in both noise-free and noise-perturbed landscapes}
\label{angleDistribution}

We now study the effect of varying the underlying angular distribution for protrusions, $f(\theta)$, and its relation to the optimal number and length of protrusions extended. 
In many migrating cell types, the biochemical and mechanical processes that govern actin dynamics are non-uniform around the cell membrane and are biased towards the front by polarity signalling pathways (for example, localised PI3K activation \cite{huang2013excitable} or Rac‐driven Arp2/3 recruitment \cite{dang2013inhibitory}). 
This forward bias in protrusion orientation, whether arising from pre‐existing polarity cues, membrane tension gradients, or feedback from nascent adhesions, can dramatically reshape the optimal trade‐off between protrusion number and size. 
In shallow or noisy gradients, a strongly front‐focused strategy may be too rigid, causing the cell to overlook off‐axis cues that would otherwise steer it back on course. 
In contrast, in steep, well-defined gradients, a tight angular focus minimises wasted energy on poorly-oriented protrusions, allowing the cell to invest energy in fewer, longer extensions and thereby move more efficiently. 

In order to study these effects within the model, we begin by considering

\begin{equation}
\label{eq:vonmises}
f(\theta;\kappa)
= \frac{\exp\bigl(\kappa\cos(\theta)\bigr)}
       {\displaystyle \int_{0}^{\pi}\exp\bigl(\kappa\cos(\phi)\bigr)\,\rm{d}\phi}
\coloneq \frac{\exp\bigl(\kappa\cos(\theta)\bigr)}{\pi\,I_{0}(\kappa)},
\qquad \theta\in[0,\pi],
\end{equation}

such that the angles at which protrusions are generated are sampled independently from a von Mises distribution supported on $[0,\pi]$, with $\kappa$ determining the bias with which protrusions are formed in the positive $x$ direction ($\theta=0$, the direction of the positive chemoattractant gradient in the model). 

By fixing $\alpha_\text{a}$, $\alpha_{\text{m}}$, $\beta$,  $g(\beta, \phi)$, and $d$ in Equation \eqref{DTER}, maximisation of the DTER for a given $n_\text{p}$ and $l_\text{p}$ simply means the maximisation of 

\begin{equation}
    \int_{0}^{\theta_{\text{crit}}} f(\theta^{''}, \kappa)\cos(\theta^{''}) \left(1 - \int_{0}^{\theta^{''}}f(\theta^{'}, \kappa)\text{d}\theta^{'}\right)^{n_\text{p}-1}\text{d}\theta^{''},
    \label{numerator}
\end{equation}

with respect to $\kappa$. In chemoattractant profiles with a noise-free gradient in the $x$ direction ($c(x,y) = c_0x/L$), it is trivial to see that increasing $\kappa$ increases the DTER monotonically as, in this idealised situation, efficient movement simply means the minimisation of $\theta_{\text{min}}$ (such that the largest possible gradient is detected, and movement is optimised to occur in the direction of the gradient). 

For migration \textit{in vivo}, a more pertinent question is whether the same relationship holds in more realistic chemoattractant profiles where the chemoattractant landscape is perturbed with spatially-correlated noise. In order to probe how environmental heterogeneity alters the optimal angular bias, we superimpose a spatially‐correlated noise field \(\eta(x,y)\) onto the idealised gradient.  Specifically, we define
\begin{equation}
      c(x,y) \;=\; \frac{c_0}{L}\,x \;+\; \sigma\,\eta(x,y),
      \label{perturbedField}
\end{equation}

where \(\eta\) is drawn from a Gaussian random field with zero mean, unit variance, and squared-exponential covariance
\[
  \mathbb{E}\bigl[\eta(\mathbf{r})\,\eta(\mathbf{r}')\bigr]
  = \exp\!\Bigl(-\|\mathbf{r}-\mathbf{r}'\|^2/4\xi^2\Bigr),
\]
where $\mathbf{r} = \mathbf{r}(x,y)^{\top}$ and $\xi$ is a parameter that determines the correlation length of the noise field.  The noise amplitude \(\sigma\) therefore sets the signal‐to‐noise ratio (SNR) of the local gradient cues. In order to study the chemoattractant landscape in Equation \eqref{perturbedField}, we now re-cast the DTER model as a stochastic agent-based model (ABM). 

We simulate cell migration in a noisy chemoattractant landscape using an ABM with discrete protrusive sampling.  The two–dimensional concentration field is defined on a square domain of side length \(L\), with $\eta(x,y)$ and $\sigma$ as defined above. The field is generated by sampling white noise on an \(N\times N\) grid, convolving with a Gaussian kernel of width \(\xi\), rescaling to standard deviation \(\sigma\), and adding the linear gradient of slope \(c_{0}/L\). 

At each discrete time, \(t\), the cell is treated as a point particle at position \(\mathbf{X}_t\) and draws \(n_\text{p}\) protrusion angles \(\{\theta_i\}\) from a von Mises distribution $f(\theta;\kappa)$ centred on the positive $x$-direction, \(\theta=0\). For each protrusion, $i$, at time $t$, the tip location is computed as
\[
\mathbf{X}_t^{(i)}=\mathbf{X}_t+l_{\text{p}}[\cos(\theta_i),\sin(\theta_i)]^T,
\]
and the concentration difference as \(\Delta c_i=c(\mathbf{X}_t^{(i)})-c(\mathbf{X}_t)\). A detection threshold 
\(\Delta c_{\rm crit}=(c_0/L)\,d_{\rm crit}\) 
is imposed, and the set \(\mathcal{M}=\{\,i:\Delta c_i\ge\Delta c_{\rm crit}\}\) is formed. If \(\mathcal{M}\) is empty, the cell moves a distance $d$ in a random direction, otherwise it selects \(i^*=\arg\max_{i\in\mathcal{M}}\Delta c_i\) and moves according to
\[
\mathbf{X}_{t+1}=\mathbf{X}_t+d[\cos(\theta_{i^*}),\sin(\theta_{i^*})]^T.
\]
After \(T\) steps the net displacement \(\Delta \mathbf{X}=\mathbf{X_T}-\mathbf{X_0}\) is recorded, and over \(M\) independent realisations of the landscape and trajectory we compute the mean of \(\Delta \mathbf{X}\). By fixing $n_\text{p}$, $l_{\text{p}}$, $\beta$, and $d_{\rm crit}$ (and hence, the energy expended in environmental sampling and movement), we vary $\kappa$, $\sigma$, and $\xi$ to quantify how protrusion directional bias influences chemotactic efficiency for a range of background environmental noises and correlation lengths. 

\begin{figure}
    \centering
    \includegraphics[width=\linewidth]{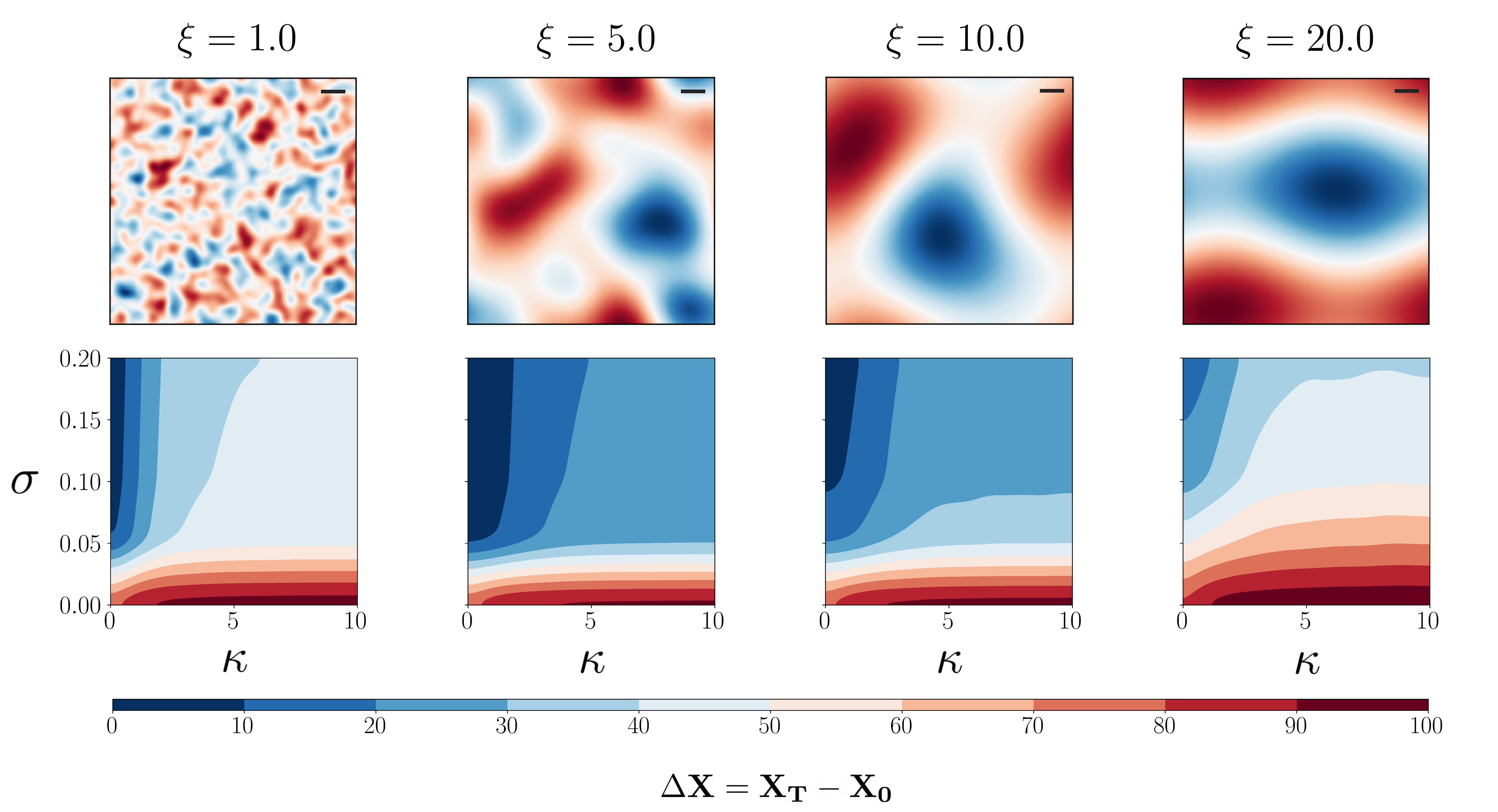}
    \caption{Heatmaps of mean displacement \(\langle \Delta \mathbf{X}\rangle\) as a function of environmental noise \(\sigma\) and protrusion orientation bias \(\kappa\), for correlation parameters \(\xi=1.0,\,5.0,\,10.0\) and \(20.0\). Each panel shows the mean final \(x\)-displacement after 100 environmental samples and subsequent movement  and \(M=10^4\) agent realisations, interpolated on a \(200\times200\) grid and smoothed with a Gaussian filter. Above each plot is a typical noise landscape, $\eta(x,y)$. At the top right corner of each landscape, the protrusion length in simulations, $l_{\text{p}}=5$, is indicated by a black line. In all simulations, $n_\text{p} = l_\text{p} = 5$, and $d=1$.}
    \label{fig:biasResults}
\end{figure}

The analysis in Figure \ref{fig:biasResults} shows that increasing the bias parameter \(\kappa\) (i.e.\ imposing a stronger front‐focused bias on protrusion orientation) yields a monotonic improvement in chemotactic efficiency in noise-perturbed chemical landscapes. By concentrating sampling in the direction of the global concentration gradient, cells reduce energy wasted on off‐axis protrusions and maximise net forward displacement per unit energy.  

However, as the characteristic length-scale of spatially correlated noise is increased, the benefit of angular bias is somewhat attenuated. When the noise correlation length (i.e. the distance over which correlations are reduced by a factor of $e$), \(2\xi\), is much smaller than the protrusion length, \(l_{\text{p}}=5\), local fluctuations decorrelate over distances shorter than the sampling scale, such that spurious peaks and troughs in the chemical concentration average out across tips and the underlying linear gradient dominates.  In this regime, a strong front‐focused bias (\(\kappa\gg1\)) yields maximal chemotactic efficiency.  As the correlation length grows to be comparable to \(l_{\text{p}}\), however, the noise becomes coherent across neighbouring protrusions---multiple tips sample the same short-ranged gradient, which can mask or reverse the true global chemoattractant gradient. This reduces the net benefit of angular bias and can even make intermediate bias optimal.  
Finally, for sufficiently large correlation lengths, the noise field appears nearly constant over each cell’s sampling region, merely shifting the concentration baseline without creating misleading local gradients, so that again a strong angular bias best leverages the true signal. 

\textit{In vivo}, chemotactic landscapes are rarely perfectly smooth. Local fluctuations arise from pulsatile secretion of chemoattractants \cite{vinet2014initiation}, variable diffusion barriers in the ECM \cite{henke2020extracellular}, and competing signals from neighbouring cells \cite{foxman1999integrating}. Our results suggest that immune cells or amoebae probing environments with very fine‐scale heterogeneities (e.g.\ interstitial tissues with dense ECM pores much smaller than their pseudopod length) will effectively filter out this high‐frequency noise and exploit strong front‐focused polarity to navigate efficiently.  Conversely, in tissues where chemoattractant gradients are distorted over length scales comparable to that of a protrusion, such as in inflamed or tumour‐remodelled stroma with tangled fibre networks \cite{provenzano2006collagen}, excessive polarisation may lead cells astray by over‐committing to spurious local peaks.  Finally, when heterogeneities occur on much larger scales (for example, across organ‐scale morphogen gradients \cite{eldar2003self}), the noise simply shifts the overall baseline and does not create noise at the length-scale of individual protrusions, so that robust polarity once again becomes the optimal strategy. These findings imply that cells may adapt their degree of polarity dynamically by down‐regulating bias when local fluctuations match their sensing radius and up‐regulating it in very smooth or very coarse environments, in order to maintain reliable chemotaxis across heterogeneous tissues.  

In summary, our analysis highlights that optimal protrusive polarity is not a fixed trait but a context‐dependent parameter. Cells must calibrate their angular bias to the reliability of chemoattractant signals. Incorporating mechanisms for feedback‐mediated adjustment of \(\kappa\) in response to measured gradient variance would be a natural extension to the model formulated here, allowing for an exploration of how real cells achieve robust chemotaxis across diverse, noisy environments.  

\subsection{Persistent movement frees cells from local concentration minima and maxima in migration}
\label{persistentMovement}

Cellular persistence refers to the continuous maintenance of a dominant migratory axis across repeated protrusion–retraction cycles and is a well-documented feature of motile cell populations \cite{petrie2009random}. Mechanistically, persistence emerges from interconnected positive-feedback loops involving actin polymerisation, adhesion dynamics, and signalling restricted to the leading edge \cite{costa2010dissection, campa2015crossroads}. Sustained pseudopod extension at the front of the cell is supported by Rac1-driven activation of the Arp2/3 complex, while PI3K-dependent PIP$_3$ accumulation further consolidates front-directed polarity. Stabilisation of protrusions and suppression of lateral fluctuations are provided by the maturation of adhesive contacts.

This inherent directional memory may allow cells to bypass misleading local fluctuations in chemoattractant concentration. Furthermore, it may also reduce the energetic demands of frequent environmental sampling in long-distance migration. By maintaining a biased protrusive front, transient adverse gradient cues that obscure global chemical landscapes cannot easily deflect cell trajectories, enabling navigation through small-scale obstacles and the escape of local chemoattractant minima and maxima. \textit{In vivo}, such persistence is essential for leukocytes squeezing through dense interstitial matrices \cite{lammermann2008rapid}, as well as for developmental cohorts such as cranial neural crest cells that must traverse heterogeneous embryonic tissues in a coherent stream \cite{kulesa2010cranial}.

In this section, we build on this biological foundation to quantify how persistence interacts with our protrusion-based sensing model, increasing energetic efficiency in noise-perturbed chemoattractant profiles. To probe how the frequency of environmental sensing influences chemotactic efficiency under varying noise amplitudes and correlation scales, we extend our agent-based simulations to systematically sweep over the sampling persistence parameter, $d$, which governs how frequently cells sample their environments with protrusions between successive movement steps. We once again vary the level of spatial heterogeneity in the chemoattractant landscape, quantified by the noise amplitude, \(\sigma\), and the noise correlation length parameter, \(\xi\). For each parameter combination, we then simulate the ABM for various values of \(d\), which in turn allows us to analyse the evolution of the DTER with respect to the level of persistence in cell movement. At each discrete time point, the cell, again approximated as a point particle, either re‐samples the environment with \(n_{\rm p}=5\) protrusions, each of length $l_\text{p}=5$, from a uniform distribution, if $d$ steps have been taken, or moves in its previously chosen direction. Gradient detection logic and parameters are as outlined in the previous section. For each triplet, \((\sigma,\xi,d)\), we execute \(M=10^4\) independent realisations of the noisy landscape and cell trajectory, each of duration \(N=100\) steps and $N/d$ environmental samples, recording the final net \(x\)-displacement, $\mathbf{X}$. By computing the sample mean, \(\langle \Delta \mathbf{X}\rangle\), across replicates, we quantify the directional bias of migration under different sensing cadences. Then, quantifying the energy expended by cells in each replicate, according to Equation \eqref{totalEnergy}, we compute the average DTER for movement. 

\begin{figure}[ht!]
    \centering
    \includegraphics[width=\linewidth]{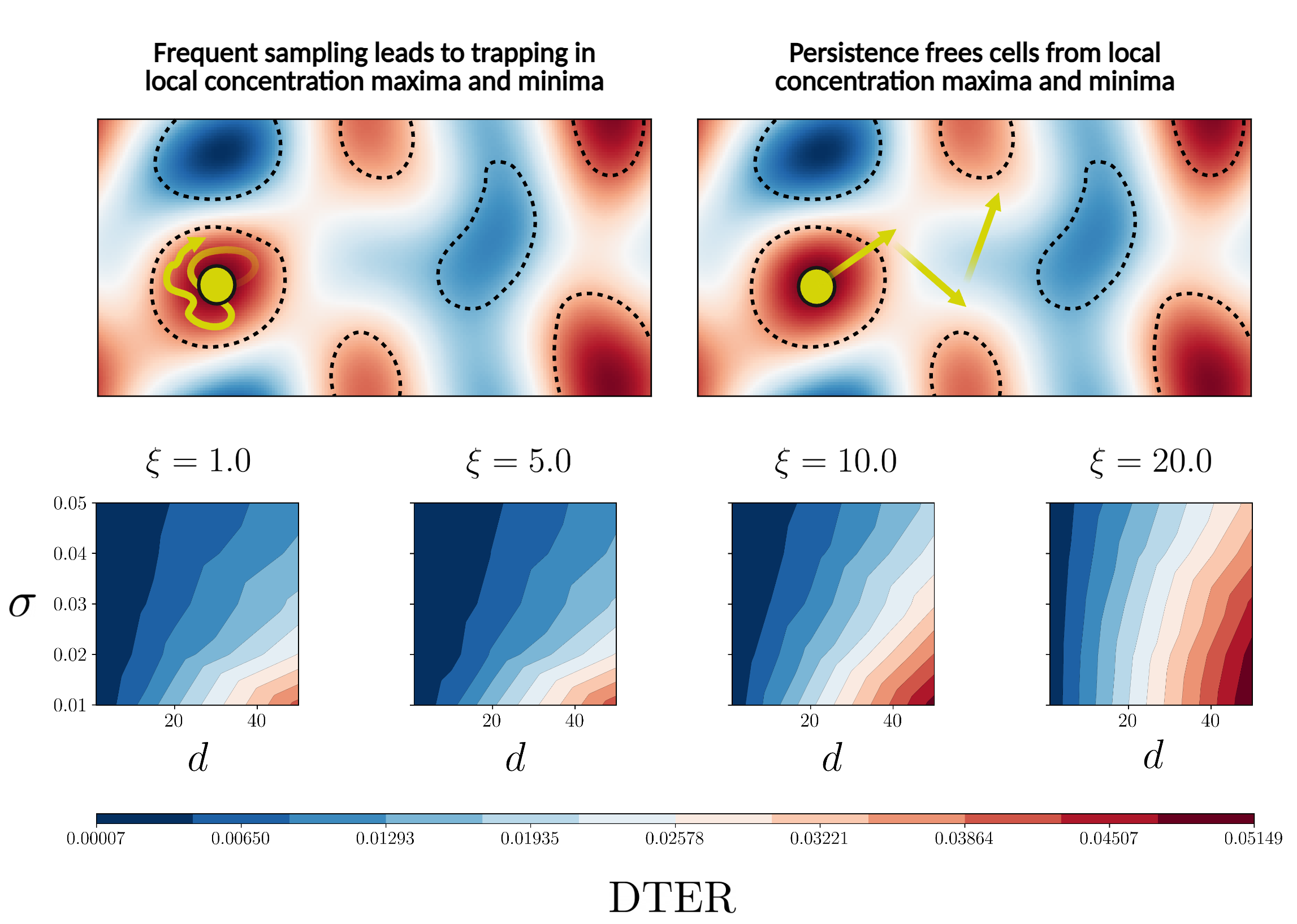}
    \caption{Heatmaps of the DTER, plotted against environmental noise amplitude $\sigma$ (vertical axis) and sensing interval $d$ (horizontal axis). Panels correspond to noise correlation length parameters $\xi=1.0,\,5.0,\,10.0,$ and $20.0$. Results are averaged over $M=10^4$ replicates, and parameters for energy expenditure and gradient detection are fixed at $\alpha_{\text{a}} = \alpha_{\text{m}} = 1$, $\beta=10^2, k=10^2/\sqrt{2}$, $\phi=1$, and $d_{\text{crit}} = 0.5$. Panels above the heatmaps illustrate how persistence in movement can free cells from local maxima and minima in chemoattractant concentration.}
    \label{fig:persistencePlots}
\end{figure}

As shown in Figure~\ref{fig:persistencePlots}, increasing the persistence interval \(d\) yields a substantial and monotonic improvement in the DTER across all noise amplitudes, \(\sigma\), and correlation length parameters, \(\xi\).  This effect increases drastically as \(\xi\) is increased. The reasons for this are two-fold. In landscapes where noise is correlated over much larger length-scales than the protrusions extended by cells, it is possible for cells that sample local chemical gradients frequently to become trapped in local maxima or minima in chemoattractant concentrations (Figure~\ref{fig:persistencePlots}). However, by moving persistently, cells are able to explore a larger proportion of the chemical landscape, and hence, are freed from local extrema in concentration. Furthermore, persistent movement means that cells expend less energy in sampling their environment with protrusions, such that the DTER is further increased.

These findings highlight persistence as a dynamically tunable strategy in chemotaxis rather than a fixed cellular trait.  \textit{In vivo}, motile cells may optimise chemotactic navigation by adjusting their sampling–movement cadence in response to local signal noise.  Up‐regulating persistence in noisy, fluctuating environments conserves energy and allows cells to explore a larger proportion of the global chemoattractant landscape. Consequently, feedback mechanisms that couple measured gradient variance to persistence may be a fundamental biological mechanism for robust, energy‐efficient chemotactic migration.  

\section{Discussion}

In this work, we developed a novel theoretical framework that frames actin-driven protrusion dynamics in eukaryotic chemotaxis as an energetic optimisation problem, wherein cells balance the benefits of chemical gradient detection against the energetic costs of protrusion extension and subsequent movement. 
Our model successfully reproduces experimentally observed morphological behaviours adopted by cells across diverse biological contexts, offering mechanistic insights into how cells may act to optimise protrusive activity under varying environmental constraints.

Our findings underscore the importance of energy-based trade-offs as a fundamental organising principle governing cellular protrusive behaviours. 
By explicitly linking protrusion length, number, substrate stiffness, and gradient detection thresholds within an energy optimisation framework, our model provides a means to predict cellular responses to environmental perturbations. 
Future experimental validation of these predictions, such as systematic variation of chemoattractant gradient steepness, controlled manipulation of substrate properties, and a study of persistence in noisy chemical landscapes could further refine the predictive power of the model and validate its broader biological applicability.

An important next step involves integrating additional biological realism into the model, such as dynamic protrusion formation and retraction, the consideration of unequal protrusion lengths, and biochemical feedback loops. 
Such enhancements would allow exploration of transient dynamics in chemotactic sensing and reveal how feedback mechanisms interact with energetic constraints to shape protrusive behaviour over biologically relevant timescales.

In conclusion, our work provides a novel theoretical lens through which to interpret and predict eukaryotic cell behaviours in chemotaxis. By elucidating how cells balance energetic constraints with information acquisition in complex environments, we contribute to a deeper understanding of cell migration, potentially informing therapeutic strategies targeting pathological migration in processes such as cancer metastasis and immune cell infiltration.

\subsubsection*{Conflict of interest statement} 
The authors declare no conflict of interest.

\subsubsection*{Authors' contributions}
\textbf{Samuel Johnson:} Conceptualisation (lead), data curation (lead); formal analysis (lead); methodology (lead); software (lead); writing–original draft (lead); writing–review and editing (supporting) \textbf{Maddy Parsons}: Writing–original draft (supporting);  writing–review and editing (lead). \textbf{Ruth E. Baker}:  Conceptualisation (supporting), supervision (lead);  writing–review and editing (lead). \textbf{Philip K. Maini:}  Conceptualisation (supporting), supervision (lead);  writing–review and editing (lead).

\subsubsection*{Acknowledgments}
S.W.S.J. receives support from the Biotechnology and Biological Sciences Research Council (BBSRC) (grant number BB/T008784/1). R.E.B. is supported by a grant from the Simons Foundation (MP-SIP-00001828). For the purpose of open access, the authors have applied a CC BY public copyright licence to any author accepted manuscript arising from this submission. The authors would like to acknowledge the use of BioRender in the production of figures. 

\bibliography{BIBL}

\appendix 

\section*{Appendix A: Model parameter sweep}

\begin{figure}
    \centering
    \includegraphics[width=0.95\linewidth]{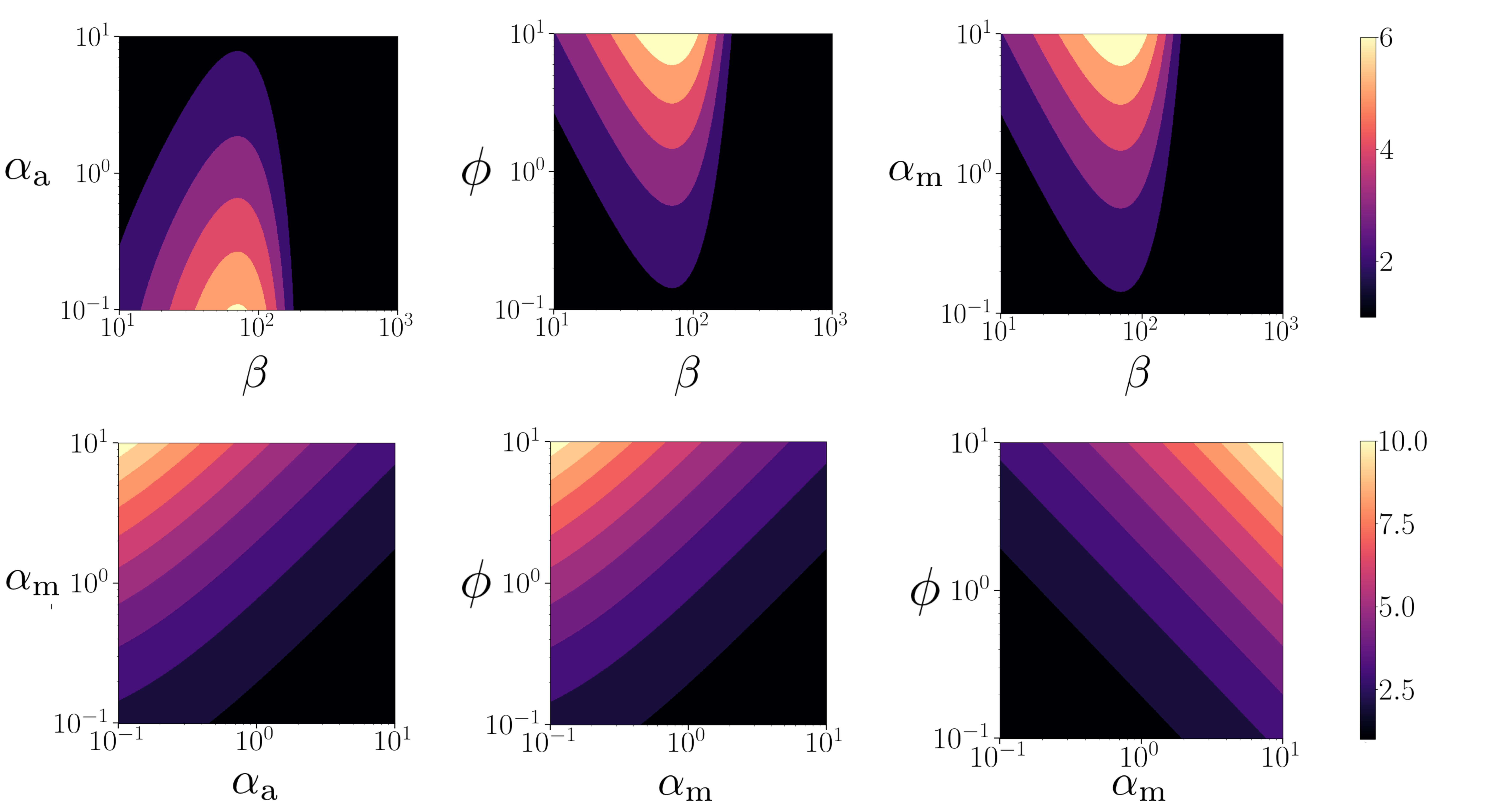}
    \caption{ $n_{\text{p}}^{\text{opt}}$ as a function of various parameter pairings for the base case of $\alpha_{\text{a}} = \alpha_{\text{m}} = \phi = 1$, $\beta=10^2$, and $k=10^2/\sqrt{2}$, with $d_{\text{crit}} = 0.5$. Heatmaps show $n_{\text{p}}^{\text{opt}}$ when two of these model parameters are varied. In general, increasing the cost of membrane deformation decreases $n_{\text{p}}^{\text{opt}}$, whereas an increase in the cost of movement, or the distance moved in between the extension of protrusions, increases $n_{\text{p}}^{\text{opt}}$. At all parameter combinations, $l_{\text{p}}$ remains at the lowest integer value above the threshold distance for detection ($d_{\text{crit}}=0.5, l_{\text{p}} = 1$).}
    \label{fig:paramSweepSI}
\end{figure}

Variation the parameters of the model for $f(\theta) = \mathrm{U}[0, \pi]$ and 

\begin{equation}
    g(\beta, \phi) = \frac{\phi\beta}{k}\exp\left(-\left(\frac{\beta}{k}\right)^{2}\right),
\end{equation}
\\
\noindent shows that varying energetic parameters leads to a variation in $n_{\text{p}}^{\text{opt}}$ for a given $d_{\text{crit}}$ (Figure \ref{fig:paramSweepSI}). Generally, we find that $n_{\text{p}}^{\text{opt}}$ increases when cells become more motile ($\phi$ is increased) or the energetic cost of movement is increased. Conversely, we find that increasing the energetic cost associated with membrane deformation $\alpha_{\text{a}}$ leads to a lower value of  $n_{\text{p}}^{\text{opt}}$. These trends are intuitive, revealing a tension between the energetic costs of membrane deformation and the energetic costs of movement --- if movement cost is high relative to deformation cost, then it is sensible to sample the environment in many directions to optimise the direction of movement and hence, maximise the DTER. Conversely, if membrane deformation costs are high relative to those associated with movement, then it is sensible to sample less, and move in potentially sub-optimal directions without spending vast amounts of energy on environmental sampling. In general, we find that over all parameter values, $l_{\text{p}}^{\text{opt}}$ remains at the lowest integer value that exceeds the threshold distance for gradient detection, $d_\text{crit}$. 

\end{document}